\documentclass[aps,prb,amsmath,twocolumn,amssymb,floatfixng,showpacs,
superscriptaddress,footinbib,multicol]{revtex4-1}
\usepackage{graphicx}
\usepackage{epstopdf}
\usepackage{dcolumn}
\usepackage{bm}
\usepackage{color}
\usepackage{amsmath}
\usepackage{braket}
\usepackage{amsfonts}
\usepackage{soul}
\usepackage{url}
\usepackage{array,tabularx}
\usepackage{graphics}
\usepackage{times}
\usepackage{subfigure}
\usepackage[export]{adjustbox}
\usepackage[colorlinks=true,linktoc=page,linkcolor=red,citecolor=blue]{hyperref}
\newcommand\x{4.4} 

\begin{document}

\title{Supersonic wave propagation in active non-Hermitian acoustic metamaterials}
\author{Kangkang Wang}
\affiliation{Key Laboratory of Modern Acoustics, MOE, Institute of Acoustics, Department of Physics, Nanjing University, Nanjing 210093, P. R. China}
\author{Felix Langfeldt}\email{f.langfeldt@soton.ac.uk}
\affiliation{Institute of Sound and Vibration Research, University of Southampton, University Road, Highfield, Southampton, SO17 1BJ,
United Kingdom}
\author{Chen Shen}
\affiliation{Department of Mechanical Engineering, Rowan University, Glassboro, NJ 08028, USA}
\author{Haishan Zou}
\affiliation{Key Laboratory of Modern Acoustics, MOE, Institute of Acoustics, Department of Physics, Nanjing University, Nanjing 210093, P. R. China}
\author{Sipei Zhao}
\affiliation{Centre for Audio, Acoustics and Vibration, Faculty of Engineering and IT, University of Technology Sydney, Ultimo, NSW 2007, Australia}
\author{Jing Lu}\email{lujing@nju.edu.cn}
\affiliation{Key Laboratory of Modern Acoustics, MOE, Institute of Acoustics, Department of Physics, Nanjing University, Nanjing 210093, P. R. China}
\author{Lea Sirota}\email{leabeilkin@tauex.tau.ac.il}
\affiliation{School of Mechanical Engineering, Tel Aviv University, Tel Aviv 69978, Israel}

\begin{abstract}

Obtaining a group velocity higher than the speed of sound in a waveguide is a challenging task in acoustic wave engineering. 
Even more challenging is to achieve this velocity increase without any intervention with the waveguide profile, such as narrowing or widening, and particularly without interfering with the passage by flexible inclusions, either passive or active. 
Here, we approach this problem by invoking concepts from non-Hermitian physics, and imposing them using active elements that are smoothly sealed within the waveguide wall. In a real-time feedback operation, the elements induce local pressure gain and loss, as well as non-local pressure integration couplings. 
We employ a dedicated balancing between the control parameters, derived from lattice theory and adjusted to the waveguide system, to drive the dynamics into a stable parity-time-symmetric regime. 
We demonstrate the accelerated propagation of a wave packet both numerically and experimentally in an air-filled waveguide and discuss the trade-off between stabilization and the achievable velocity increase.
Our work prepares the grounds for advanced forms of wave transmission in continuous media, enabled by short and long range active couplings, created via embedded real-time feedback control.

\end{abstract}

\maketitle

\maketitle

\begin{figure*}[htpb]
    \centering
    \begin{tabular}{c}
    \setlength{\tabcolsep}{6pt}
  \begin{tabular}{c c} 
  \begin{tabular}{c}
       \textbf{(a)}  \\ \includegraphics[width=7.4 cm, valign=t]{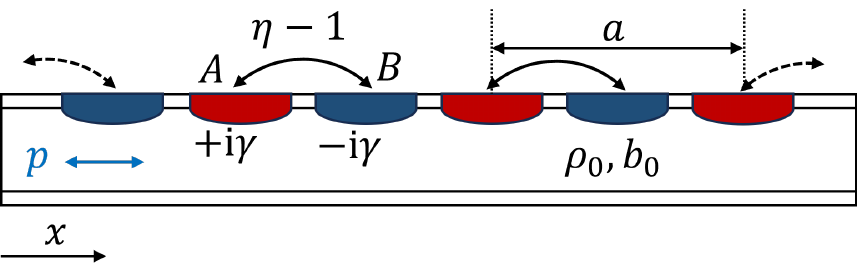}  \\ \textbf{(b)}  \\ \includegraphics[width=7.0 cm, valign=t]{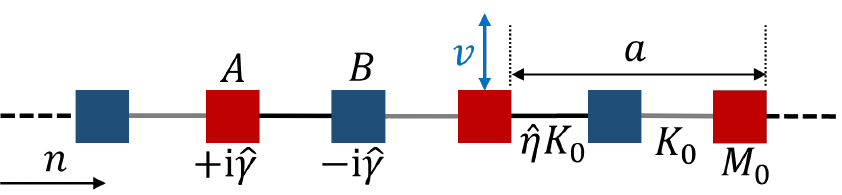} 
  \end{tabular} & \begin{tabular}{c}
      \textbf{(g)}   \\
     \qquad  \includegraphics[width=7cm]{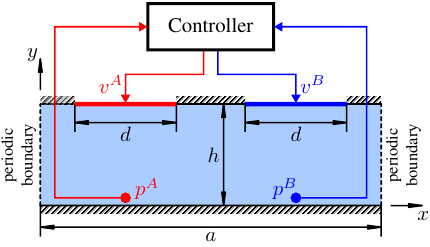}
  \end{tabular}
                  \end{tabular}             \\
          \def\arraystretch{1.4} 
           \setlength{\tabcolsep}{0pt}
          \begin{tabular}{c c c c}
          \textbf{(c)} & \textbf{(d)} & \textbf{(e)} & \textbf{(f)} \\
           \includegraphics[width=\x cm, valign=c]{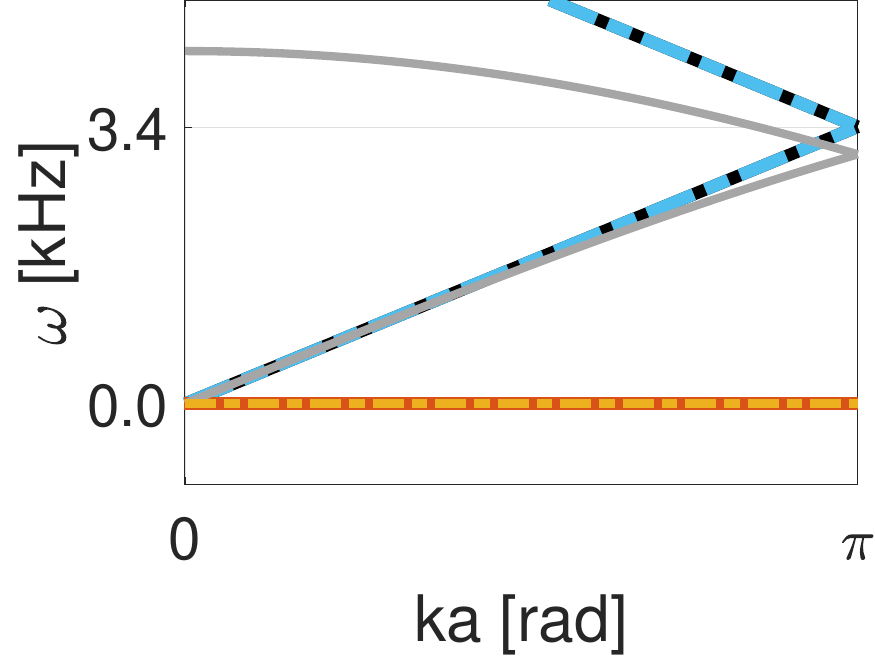}    & \includegraphics[width=\x cm, valign=c]{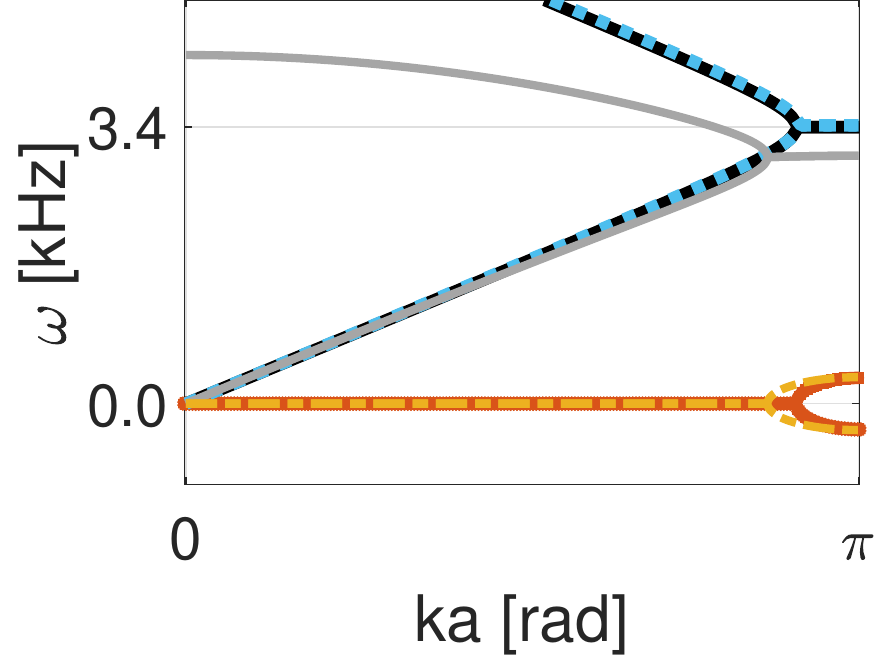}   & \includegraphics[width=\x cm, valign=c]{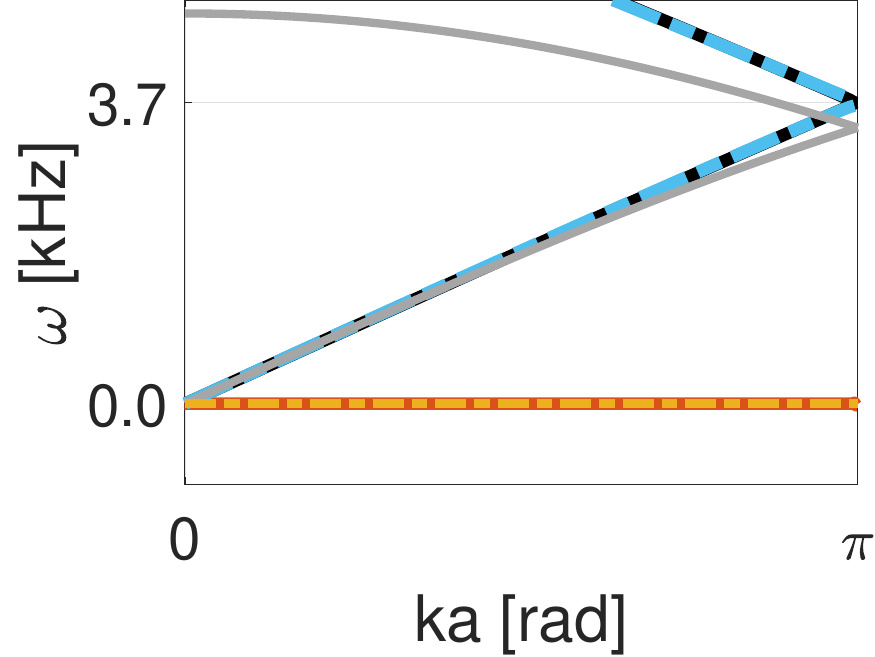}  & \includegraphics[width=\x cm, valign=c]{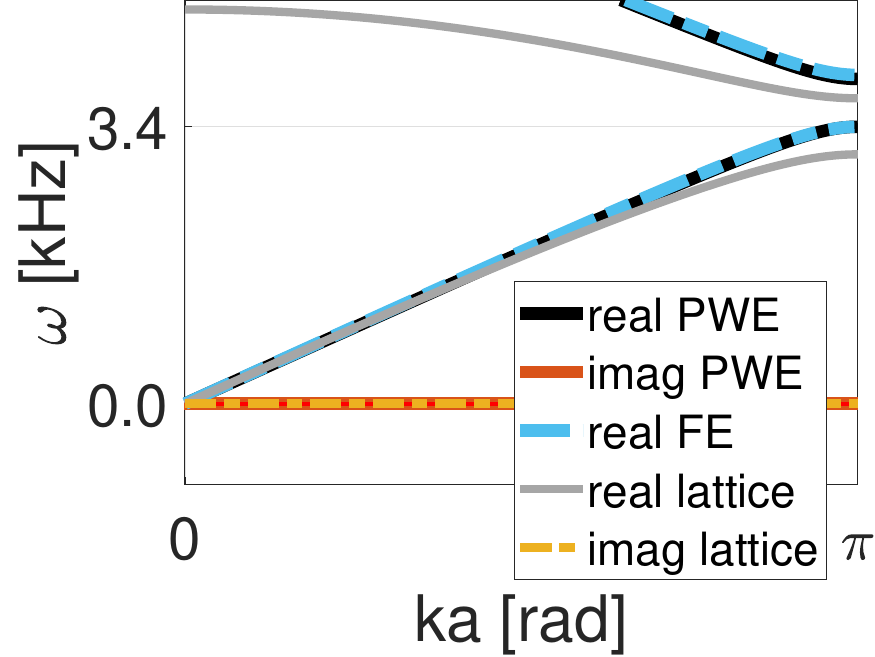} 
          \end{tabular}
    \end{tabular}
    \caption{The active acoustic metamaterial model and its spectral properties. (a) The waveguide schematics. (b) Analogous lattice schematics. (c)-(f) Frequency dispersion of the periodic system. Gray and yellow - lattice real and imaginary spectrum by Eq. \eqref{eq:Hamiltonian}. Black, orange - waveguide real and imaginary spectrum calculated via PWE by Eqs. \eqref{eq:lin_eig}-\eqref{eq:lin_eig_mat}. Cyan - waveguide real spectrum calculated via FE. $\beta=1$ was set in the waveguide. (c) Nominal stable, $\hat{\gamma}=\gamma=0$, $\hat{\eta}=\eta=1$. (d) Unstable, $\hat{\gamma}=\gamma=0.32$, $\hat{\eta}=\eta=1$. (e) PT-symmetry-restored stable, $\hat{\gamma}=0.32$, $\gamma=0.29$, $\hat{\eta}=1.5$, $\eta=2$. (f) Gapped stable, $\hat{\gamma}=\gamma=0$, $\hat{\eta}=1.5$, $\eta=2$. (g) Sketch of the FE model setup for numerically calculating the dispersion curves.}
    \label{fig:schematics_and_dispersion}
\end{figure*}

\section{Introduction}

Non-Hermitian systems, where interactions and exchange of energy with the surrounding environment are allowed, have been shown to exhibit unique properties in recent years \cite{ashida2020non,ding2022non}. Their effective non-Hermitian Hamiltonians endow exceptional properties and topologies that go beyond conventional Hermitian counterparts. A prime example is parity-time (PT) symmetry and the emergence of exceptional points (EPs), which has garnered great research interest \cite{el2018non,miri2019exceptional}. The eigenstates of these PT-symmetric systems coalesce in the parameter space, and real eigenvalues are possible with non self-adjoint Hamiltonians. Initially a purely mathematical model \cite{bender1998real,bender2007making}, the concept was later successfully tested in photonic systems with balanced loss and gain potentials thanks to the equivalence between the paraxial electromagnetic wave equation and the Schrödinger equation \cite{zhao2018parity}. Since then, numerous PT-symmetry and EP-related effects have been discovered, which suggested new functionalities and applications by tailoring the complex energy profile of the systems under study, including single-mode lasers \cite{feng2014single}, enhanced sensitivity \cite{hodaei2017enhanced}, unidirectional invisible cloaking \cite{lin2011unidirectional,sounas2015unidirectional}, and so on.

In the field of acoustics and elastodynamics, researchers have studied similar concepts and revealed a plethora of intriguing phenomena \cite{zhu2014pt,shi2016accessing,liu2018unidirectional,wang2019extremely,stojanoska2022non,cai2023absorption,huang2024acoustic}. 
Because of the absence of natural gain materials, a common approach is implementing an equivalent model with only lossy or lossless media, leaving part of the parameter space untapped. Although these systems are successful in terms of constructing complex effective Hamiltonians, recent studies have suggested that certain effects can only be induced by real gain-loss modulations \cite{xia2021nonlinear,stegmaier2021topological,li2022gain}. The use of gain media introduces external energy and broadens the utility of the parameter space. Several previous attempts have implemented real gain media using different techniques, such as electro-thermoacoustic coupling \cite{hu2021non,pernas2022theory}, background airflow \cite{auregan2017pt}, energy injection \cite{christensen2016parity,yang2022design}, or active control elements \cite{zhang2021acoustic,wang2024broadband}. 
Nevertheless, the non-Hermitian phenomena demonstrated so far have been focused mostly on the property of nonreciprocity, and less on the control of wave propagation velocity. 
In addition, the common realizations of active couplings, in particular in acoustic waveguides, involve either alternations of the waveguide geometry, or placement of the actuators in the waveguide cross-section \cite{fleury2015invisible,zhang2021acoustic,maddi2024exact}.

In this work, we address spatially continuous media, such as acoustic waveguides, hybridized with discretely-spanned active elements, which are seamlessly embedded in the waveguide wall. 
We program these elements to actively control the wave propagation velocity in a plain waveguide with a uniform cross-section. This is useful for applications for which the passage of fluid through the waveguide cannot be blocked.
Motivated by control schemes for purely discrete PT-symmetric media, such as lattices \cite{szameit2011pt,benisty2024controlled}, we derive the required couplings to speed-up wave packet propagation in the waveguide.
Specifically, the active units control the onsite loss and gain profiles, as well as the couplings between the units, to create an effective PT-symmetric system.
Utilizing the concept of feedback-based media \cite{hofmann2019chiral,rosa2020dynamics,helbig2020generalized,jana2023gravitational,zhu2023higher,langfeldt2023controlling,wen2023acoustic,halder2024circuit,maddi2024exact,jana2025invisible}, we realize these couplings in a real-time closed loop process.

At the first stage, we design a theoretical model to describe the controlled waveguide. Then, we suggest an analogous discrete model of a mass-spring lattice, and derive the relation between the control parameters that lead to a group velocity increase in the lattice, and in the same time guarantee its dynamical stability. 
Mapping back to the waveguide system, we obtain the required control parameters therein. 
Despite the differences between purely discrete and hybrid continuous-discrete systems (e.g., due to time delays or near field effects), we show that a faster group velocity compared to the background medium is possible in the waveguide system with judiciously tailored feedback modulation profiles.

Owing to the central role that stability plays in actively controlled wave systems \cite{kovacevich2024stability}, we derive the theoretical criteria for stable wave dynamics in the metamaterial so that the wave amplitude is not growing during the propagation. 
We confirm our theoretical predictions of obtaining a group velocity higher than the speed of sound in air by carrying out experiments in an acoustic waveguide, and discuss the limitations of the achievable stability in the actual system.
Our results showcase the implementation of active acoustic wave control in combination with PT-symmetry to support stable faster-than-sound dynamic pulse transmission. 
The work facilitates the spatio-temporal modulation of signals with real-time tuning capabilities and may find applications in acoustic communication and more.

\section{The target waveguide model}

We consider an acoustic waveguide that supports propagation of sound pressure waves $p$ in a fluid of mass density $\rho_0$ and bulk modulus $b_0$, as illustrated in Fig. \ref{fig:schematics_and_dispersion}(a). Active elements are connected to the waveguide wall in a periodic spacing $a/2$, positioned at $x_n$, and facing inwards. 
These elements produce acoustic control velocities $v_n$, and are incorporated in the field equation as
\begin{equation}  \label{eq:cont_OL}
    \frac{1}{c^2}p_{tt}(x,t)=p_{xx}(x,t)+\rho_0 \beta\sum_n \dot{v}_n(t)\delta(x-x_n),
\end{equation}
where $\delta(\cdot)$ is the Dirac delta function, and $\beta$ is the ratio between the active element area $S_d$ and the waveguide cross-section $S_w$. 
The actively controlled waveguide constitutes a hybrid continuous-discrete medium.
The role of the inputs $v_n$ is to increase the group velocity inside this medium, $v_g$, beyond the background speed of sound $c$. 
Motivated by control schemes for PT-symmetric lattices~\cite{szameit2011pt,benisty2024controlled}, we define $A$ and $B$ sites alternating with the periodicity of $a$. Using feedback control loops, the elements at these sites respectively induce local gain and loss defined by the parameter $\gamma$, as well as an additional coupling between each $A$-$B$ pair, defined by the parameter $\eta$. 
We set the control inputs at these sites to 
\begin{equation} \label{eq:A_B_velocities}
v^{A/B}_n(t)=\frac{\eta-1}{\rho_0a}\int_{0}^{t}[p^{B/A}_n(t)-p^{A/B}_n(t)]\mathrm{d}t\pm\frac{\gamma}{z_0}p^{A/B}_n(t),
\end{equation}
where $p^{A/B}_n(t)$ is a compact form of $p(x^{A/B}_n,t)$, and $z_0=\rho_0c$ is the specific acoustic impedance.
In closed loop, the pressure field in the waveguide
is governed by
\begin{equation}  \label{eq:cont}
\begin{split}
    &\frac{1}{c^2}p_{tt}(x,t)=p_{xx}(x,t) \\&+\beta\sum_n \frac{\eta-1}{a}(p^B_n(t)-p^A_n(t))\delta(x-x^A_n)\\&+\beta\sum_n \frac{\eta-1}{a}(p^A_n(t)-p^B_n(t))\delta(x-x^B_n)\\&+\beta\frac{\gamma}{c}\sum_n\left[\dot{p}^A_n(t)\delta(x-x^A_n)-\dot{p}^B_n(t)\delta(x-x^B_n)\right].
\end{split}  
\end{equation}
It is then required to derive a relation between $\gamma$ and $\eta$ so that $v_g>c$, but also that the system's stability is preserved, i.e., that the waves propagate with non-growing amplitudes.
We thus consider an auxiliary model -- an analogous mass-spring lattice, which is inherently discrete, Fig. \ref{fig:schematics_and_dispersion}(b), and derive the $\gamma-\eta$ relation for this model first. 
Each lattice site has a single degree of freedom, the vertical displacement $y$. The masses $M_0$ and spring constants $K_0$ are analogous to the bulk modulus and mass density of the fluid as $M_0=aS_w/(2b_0)$ and $K_0=2S_w/(\rho_0a)$, and the site velocity $v=\dot{y}$ maps to the pressure $p$. 
Defining $\hat{\eta} K_0$ as a spring constant different to the nominal $K_0$, and $\mp \mathrm{i} \hat{\gamma} Z_0$ as an onsite viscous damper (anti-damper) connected to ground, where $Z_0=\sqrt{M_0K_0}$ is the mechanical impedance, the dynamics of the $n$-th $A$ and $B$ sites of the dimer lattice can be formulated as
\begin{equation} \label{eq:lattice_n_eq}
    \omega_0^{-2}\Ddot{y}^{A/B}_n=y^{B/A}_{n\mp 1}+\hat{\eta} y^{B/A}_n -(1+\hat{\eta}) y^{A/B}_n\pm\hat{\gamma}\omega_0^{-1} \dot{y}^{A/B}_n,
\end{equation}
where $\omega_0=\sqrt{K_0/M_0}$. 
The mapping between the $\gamma$ and $\eta$ couplings of the waveguide, and $\hat{\gamma}$ and $\hat{\eta}$ couplings of the lattice then takes the form (App. \ref{cont_discrete_map})
\begin{equation} \label{eq:balance}
    \hat{\gamma}= \beta\gamma \quad , \quad \hat{\eta}= \beta\tfrac{1}{2}(\eta-1)+1.
\end{equation}
In order to derive the relation between $\hat{\gamma}$ and $\hat{\eta}$ for the group velocity increase, we insert the solution $\left[\mkern-8mu\begin{array}{cc} y^A_n & y^B_n \end{array}\mkern-8mu\right]^T=\left[\mkern-8mu\begin{array}{cc} \overline{y}^A & \overline{y}^B \end{array}\mkern-8mu\right]^Te^{\mathrm{i}(kna-\omega t)}$ in Eq. \eqref{eq:lattice_n_eq} , where $k$ is the wavenumber, and $\overline{y}^A,\overline{y}^B$ are the respective wave amplitudes.
Defining the normalized frequency $\Omega=\omega/\omega_0$, we obtain the augmented eigenvalue problem $\left(\Omega\textbf{I}-\mathcal{H}\right)\overline{\textbf{y}}=\textbf{0}$, with the augmented eigenvector $\overline{\textbf{y}}$ including $\overline{y}^A$, $\overline{y}^B$, and two auxiliary states. The effective $4\times 4$ Hamiltonian $\mathcal{H}$ is given by
\begin{equation}  \label{eq:Hamiltonian}
\mathcal{H}=\left(\mkern-8mu\begin{array}{cc}
       \textbf{0}  & \textbf{I} \\
       \mathcal{H}_0+(1+\hat{\eta})\textbf{I}  & \mathrm{i}\hat{\gamma}\bm{\sigma}_z
    \end{array}\mkern-8mu\right), \; \mathcal{H}_0=\left(\mkern-8mu\begin{array}{cc} 0 & f \\ f^* & 0 \end{array}\mkern-8mu\right),
\end{equation}
where $f=-\left[\hat{\eta}+e^{-\mathrm{i}ka}\right]$ and $\bm{\sigma}_z$ is the Pauli matrix. 
For $\hat{\gamma}>0$ this Hamiltonian is non-Hermitian \cite{ashida2020non}. 
The solution of the eigenvalue problem in Eq. \eqref{eq:Hamiltonian} gives the lattice dispersion relation, and consequently the required relation between $\hat{\gamma}$ and $\hat{\eta}$ for dynamical stability \cite{benisty2024controlled}. This is illustrated in Fig. \ref{fig:schematics_and_dispersion}(c)-(f). For the nominal lattice, i.e. for $\hat{\gamma}=0$ and $\hat{\eta}=1$, the spectrum is purely real, as expected, Fig. \ref{fig:schematics_and_dispersion}(c), where the top band is idle (due to the lattice constant folding to $a/2$). 
When increasing $\hat{\gamma}$, the spectrum becomes complex-valued around the band crossing point, turning it into an exceptional cut \cite{miri2019exceptional}, Fig. \ref{fig:schematics_and_dispersion}(d) (e.g. for $\hat{\gamma}=0.32$). The time response at the $n$-th node then takes the form $y_n(t)\propto e^{\omega_It}e^{\mathrm{i}(kna-\omega_Rt)}$, where $\omega_R$ and $\omega_I$ are the real and imaginary components of the frequency, respectively. The response thus grows unbounded, indicating the system's dynamical instability. 
In particular, for 
\begin{equation} \label{eq:gamma_eta_balance}
    \hat{\gamma}^*=\sqrt{2}\left(\sqrt{\hat{\eta}}-1\right)
\end{equation}
the lattice spectrum is restored to be purely real, as illustrated in Fig. \ref{fig:schematics_and_dispersion}(e) for $\hat{\eta}=1.5$ and $\hat{\gamma}=0.32$, while preserving the non-Hermiticity of the Hamiltonian in Eq. \eqref{eq:Hamiltonian}. This transition indicates the restoration of the PT-symmetric phase~\cite{bender1998real}. For any $\hat{\gamma}<\hat{\gamma}^*$ the spectrum remains real, albeit gapped, e.g. as shown in Fig. \ref{fig:schematics_and_dispersion}(f) for $\hat{\eta}=1.5$ and $\hat{\gamma}=0$, thus forming the lattice stability region in the $\hat{\gamma}-\hat{\eta}$ plane.
Remarkably, the new crossing point of the dispersion curves for $\hat{\gamma}=\hat{\gamma}^*$, which is an exceptional point, occurs at a higher frequency than for the nominal case, indicating an increase in group velocity. This increase is given by 
\begin{equation} \label{eq:vg_increase}
    v_g=v_0\sqrt{\tfrac{1}{\sqrt{2}}\hat{\gamma}^*+1},
\end{equation}
suggesting that it is possible to exceed the Hermitian group velocity $v_0$, and to guarantee dynamical stability (App. \ref{Lattice}). 
Back to the actual waveguide, we calculate the dispersion of Eq. \eqref{eq:cont} using the plane wave expansion (PWE) method \cite{chaunsali2018subwavelength}, in which a series solution of traveling harmonic waves, $p(x,t)=\mathrm{e}^{\mathrm{i}\omega t}P(x)$, is assumed. Here, $P(x)=\sum_{m=-M}^M\mathrm{e}^{-\mathrm{i}(k+m)\textbf{b}_1\cdot x\textbf{d}_1}\overline{p}_m$, $\textbf{b}_1=\frac{2\pi}{a}\hat{\textbf{e}}_1$ and $\textbf{d}_1=a\hat{\textbf{e}}_1$ respectively span the momentum and the real spaces, $\overline{p}_m$ is the pressure field amplitude, and $M$ is the approximation order (App. \ref{PWE_details}). Due to the continuity of the acoustic medium there is an infinite number of bands, where $M$ defines how many of these bands are plotted. The resulting spectrum is depicted on top of the lattice spectrum in Figs. \ref{fig:schematics_and_dispersion}(c)-(f) in low frequencies for an air -filled waveguide. The waveguide cross-sectional area is assumed equal to that of the active elements, i.e. $\beta=1$, with a spacing of $a=5$ $\mathrm{cm}$, and $M=4$. 

\begin{figure*}[htpb]
    \begin{center}
        \begin{tabular}{c c c c}
  \textbf{(a)}  &  \includegraphics[height=4 cm, valign=t]{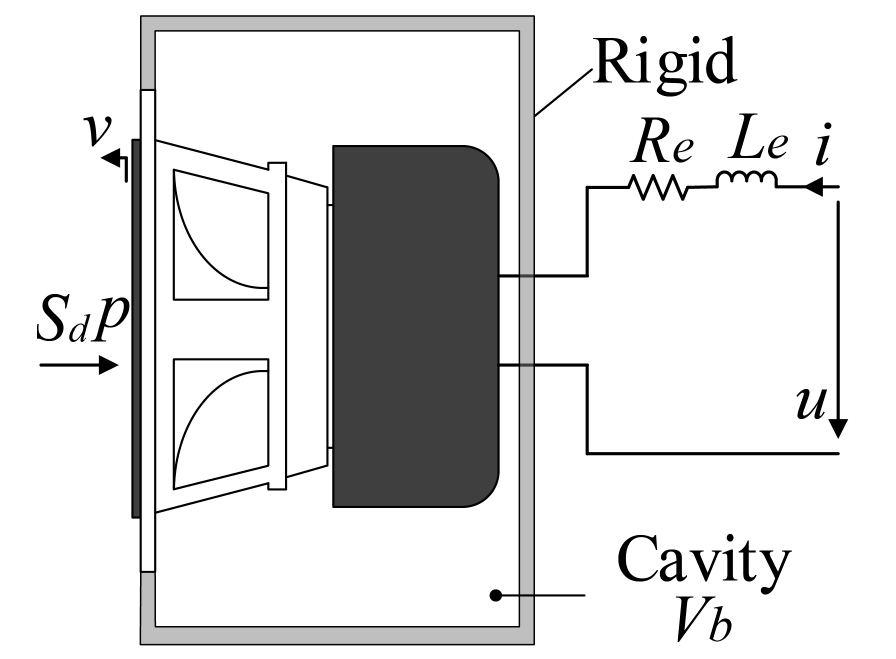}     &  \textbf{(b)} &
    \includegraphics[height=4.2 cm, valign=t]{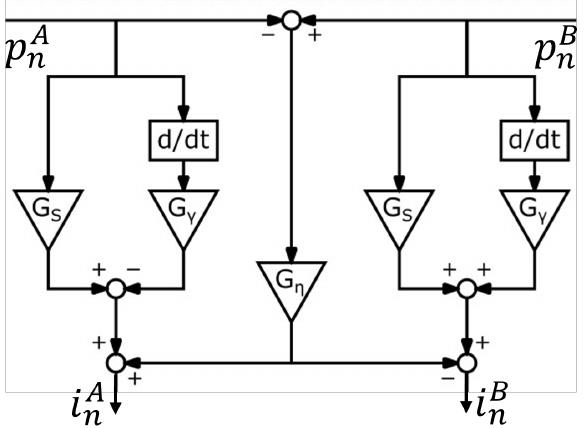}    \end{tabular}\\
    \setlength{\tabcolsep}{1pt}
    \begin{tabular}{c c c}
    \textbf{(c)}  &   \textbf{(d)}  &  \textbf{(e)}  \\ \includegraphics[height=3.5 cm, valign=b]{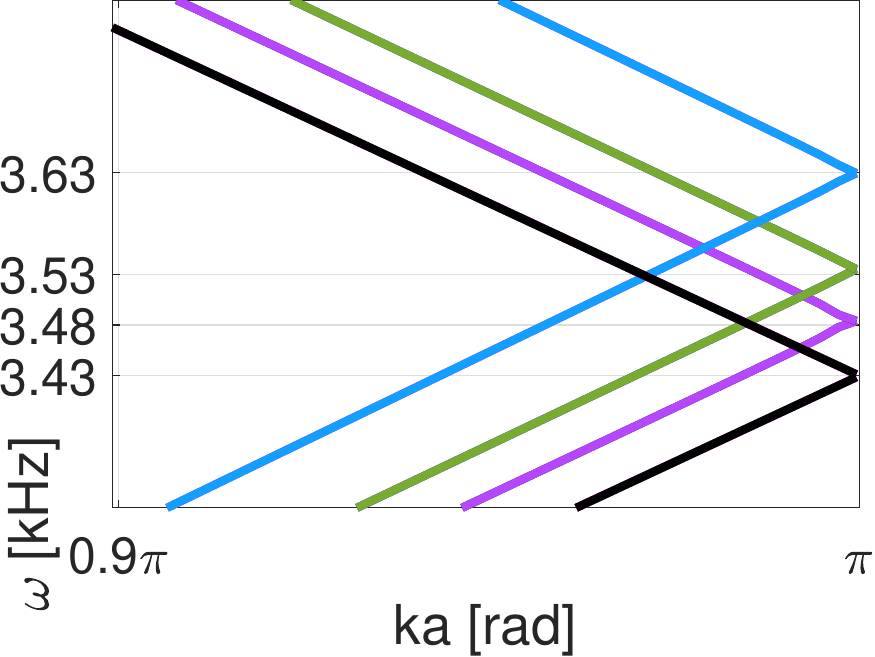}       &     \includegraphics[height=3.7 cm, valign=b]{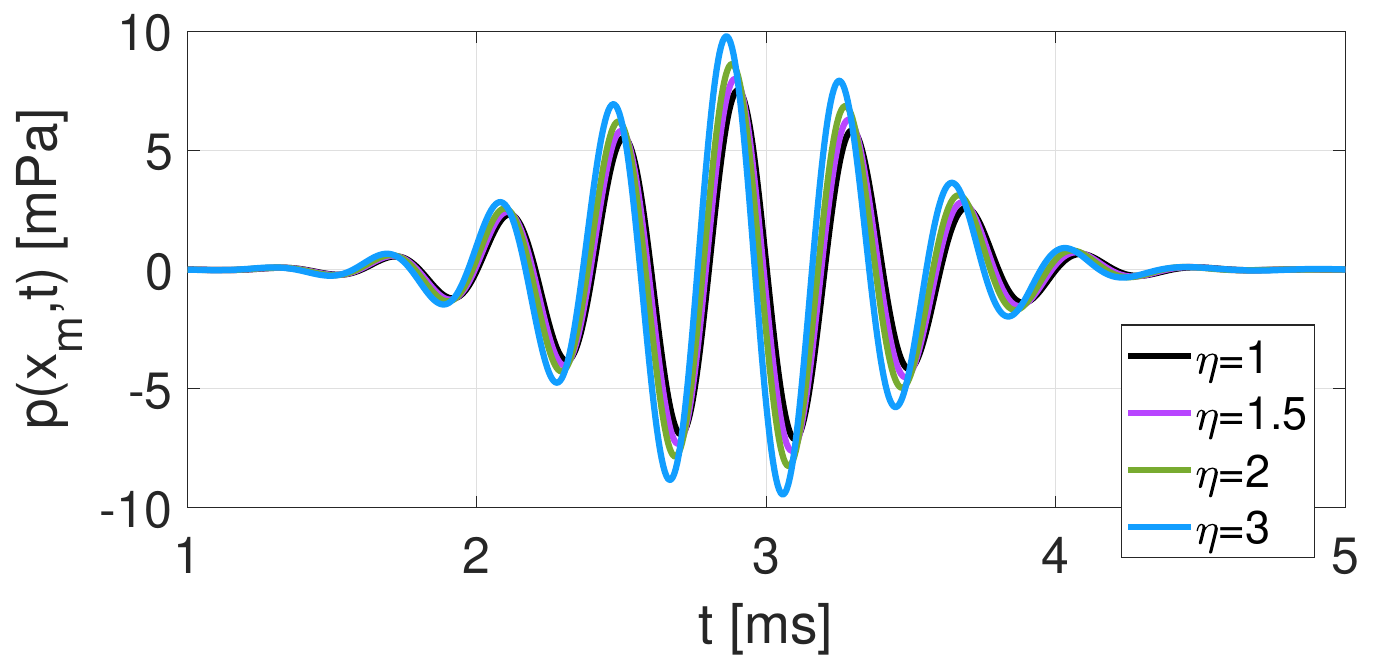}     &  \includegraphics[height=3.6 cm, valign=b]{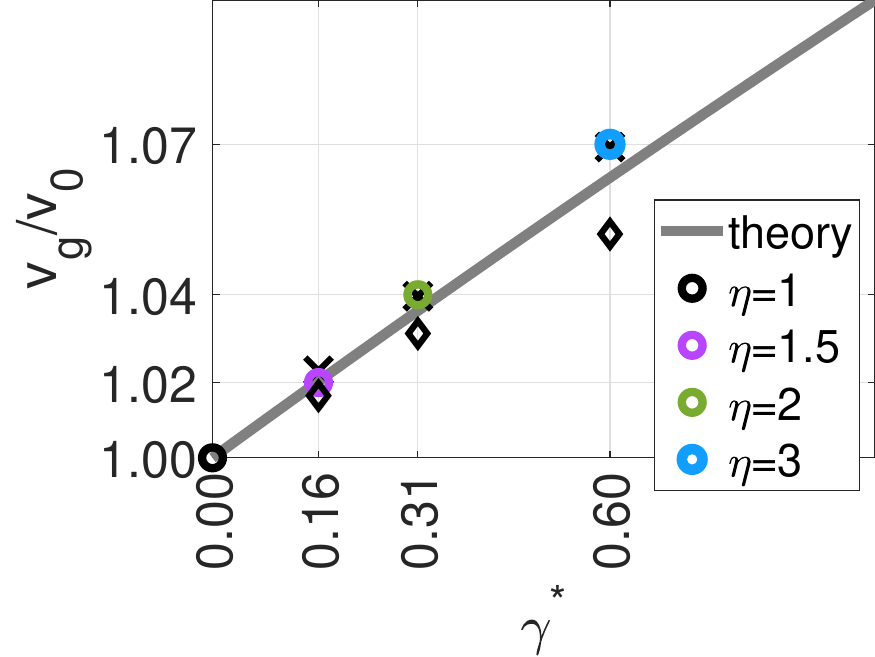}
        \end{tabular}
    \end{center}
    \caption{Waveguide realization in a feedback control setup using current-driven electroacoustic transducers. (a) Schematic of the electrodynamic speaker. (b) The controller structure. (c) The waveguide dispersion relation obtained from PWE for $a=5$ cm and $\beta=0.32$, featuring the uncontrolled case $\gamma=0,\eta=1$ (black), and the controlled cases $\gamma=0.16,\eta=1.5$ (purple), $\gamma=0.31,\eta=2$ (green), and $\gamma=0.60,\eta=3$ (blue). (d) Time domain responses to a Gaussian wavepacket centered at $\omega=2.5~\mathrm{kHz}$, calculated via FE. (e) The normalized group velocities obtained from the time domain simulations (circles), and from the corresponding dispersion plots in panel (c) ($\omega=2.5~\mathrm{kHz}-$ black diamonds, $\omega=0~\mathrm{kHz}-$ black crosses), plotted on top of the theoretical expression Eq. \eqref{eq:vg_increase}  via Eq. \eqref{eq:balance} (gray).}
    \label{fig:speaker}
\end{figure*}

Fig. \ref{fig:schematics_and_dispersion}(c) depicts the spectrum of the uncontrolled waveguide. The crossing point of the main band with its folding, which occurs at $ka=\pi$, directly implies the slope of $343$ $\mathrm{m/s}$, which, as expected, equals $c$, the speed of sound in air. 
We then begin to increase $\gamma$ while keeping $\eta=1$. Similarly to the lattice system, we substitute $\gamma=0.32$ and an imaginary spectrum appears, as depicted in Fig. \ref{fig:schematics_and_dispersion}(d). To eliminate the imaginary spectrum we increase $\eta$ as well. Since for $\hat{\gamma}^*=0.32$ the lattice balance relation in Eq. \eqref{eq:gamma_eta_balance} reads $\hat{\eta}=1.5$, the mapping to the waveguide in Eq. \eqref{eq:balance} implies $\eta=2$. 

However, as the analogy between the discrete and continuous systems is valid only in the long wavelength (low frequency) regime, a discrepancy between the underlying spectra is observed at the vicinity of the crossing point $ka=\pi$ (given by $\omega=2\sqrt{2}\sqrt{\sqrt{\hat{\eta}}}c/a$ in the lattice case). Therefore, to restore the PT-symmetry in the waveguide for a given $\eta$, the value of $\gamma$ stemming from Eqs. \eqref{eq:balance} and \eqref{eq:gamma_eta_balance} needs to be slightly modified (the particular modification value depends on $\eta$, App. \ref{cont_discrete_map}).
For $\eta=2$, the balance is then obtained for $\gamma^*=0.29$, as shown in Fig. \ref{fig:schematics_and_dispersion}(e). 
Decreasing $\gamma$ below $\gamma^*$ keeps the spectrum real, albeit gapped, as in the lattice, as illustrated in Fig. \ref{fig:schematics_and_dispersion}(f) for $\eta=2$ and $\gamma=0$. 

To confirm the results obtained via the PWE, we calculate the waveguide's frequency dispersion using finite element (FE) analysis. As shown in Fig. \ref{fig:schematics_and_dispersion}(g), the domain is modeled as two-dimensional, representing a cross-section of the waveguide with the main propagation axis $x$ and the vertical axis $y$. The height of the waveguide $h$ and the actuator length $d$ are explicitly included in the model. The actuators are represented by velocity sources, and point probes are used at the waveguide wall opposite to the actuators (similar to the experimental setup described in Sec. \ref{Experiment_sec}) to obtain the pressures $p^A$ and $p^B$. Using these pressure values, the control law given in Eq. \eqref{eq:A_B_velocities} is implemented to drive the velocity sources and create the required couplings. The ends of the waveguide section of length $a$, representing one unit cell, are terminated using periodic boundary conditions. All other boundaries are set to be sound hard. The dispersion curves were calculated by computing the eigenvalues of the periodic system with different $ka$ values. The FE results are depicted in Fig. \ref{fig:schematics_and_dispersion}(c)-(f) on top of the PWE results, nearly coinciding with each other. 

\section{Controller design}

\begin{figure*}[t]
    \centering
    \begin{tabular}{c}
   \def\arraystretch{1.4}
         \begin{tabular}{c c}  
              \textbf{(a)} &   \includegraphics[width=16.4 cm, valign=t]{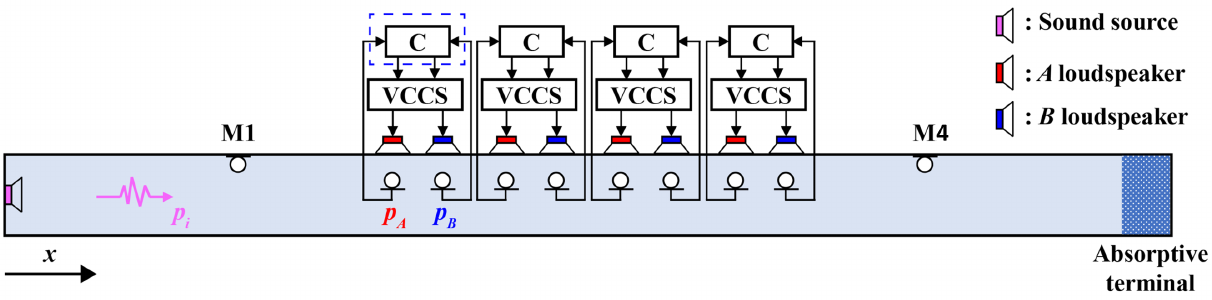} \\
              \textbf{(b)} &   \includegraphics[width=16.4 cm, valign=t]{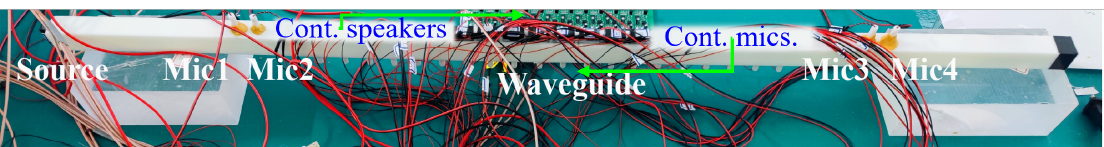}
           \end{tabular}  \\
              \setlength{\tabcolsep}{0pt}
         \begin{tabular}{c c c}         
           \textbf{(c)} &  \textbf{(d)}  & \textbf{(e)} \\
           \includegraphics[height=3.8 cm, valign=c]{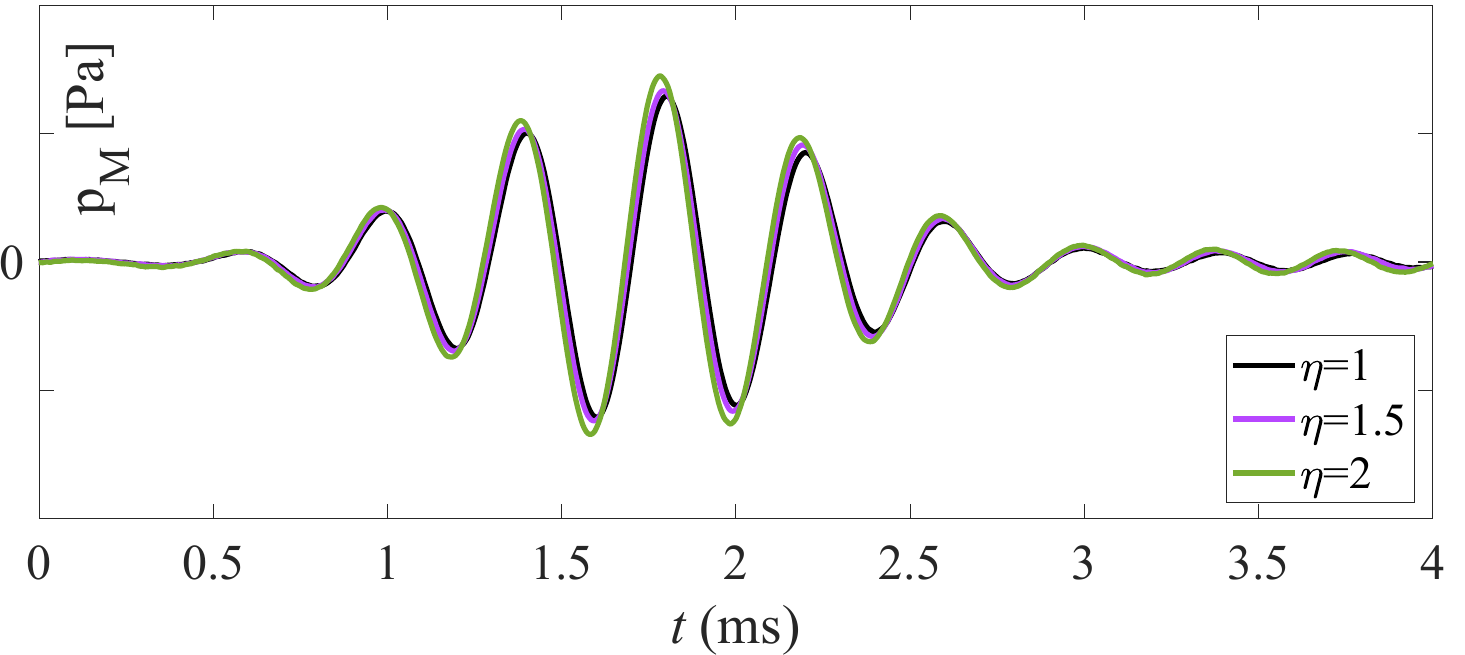}  & \includegraphics[height=4.0 cm, valign=c]{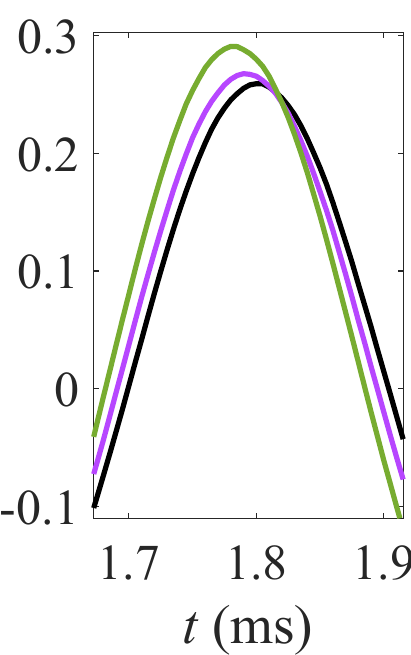} & \includegraphics[height=3.8 cm, valign=c]{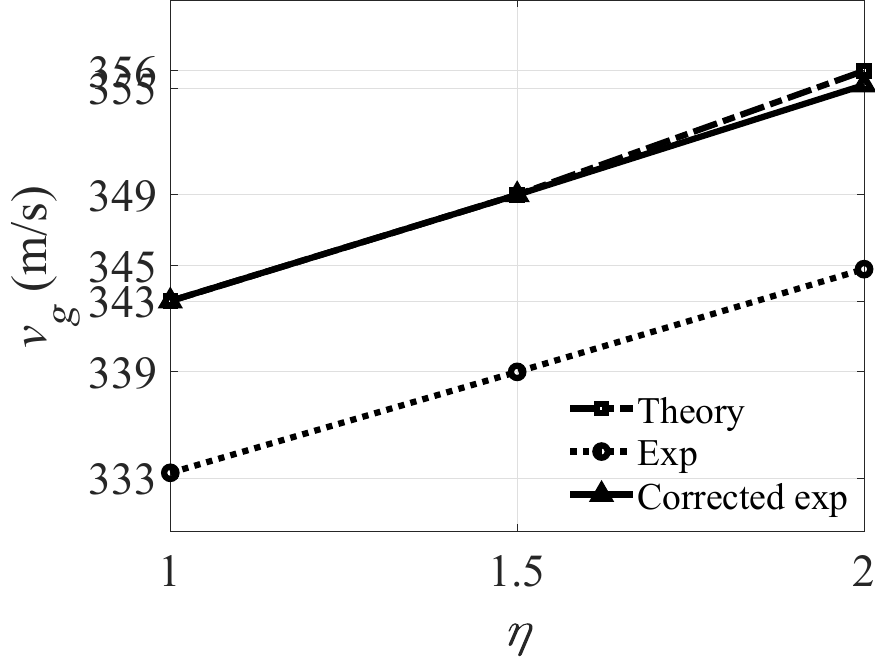}
          \end{tabular}
    \end{tabular}
    \caption{Experimental demonstration of accelerated wave packet propagation. (a) Schematic of the experimental setup. (b) Photograph of the waveguide, comprising 8 active sites.
    (c) The measured waveform at microphone Mic4 relative to microphone Mic1 as a function of time for $\eta=1$, black, $\eta=1.5$, purple, and $\eta=2$, green. (d) Zoom-in of the measured waveform in (c). (e) Comparison of the calculated group velocities with the measured results. Dashed - the theoretical values obtained from Eq.  \eqref{eq:vg_increase} via Eq. \eqref{eq:balance}. Dotted - the measured values. Solid - the measured values adjusted to the Hermitian velocity of 343 m/s.}
    \label{fig:experiment}
\end{figure*}

We realize the control velocity sources using electroacoustic transducers, which replace the ideal actuators in Fig. \ref{fig:schematics_and_dispersion}(g). 
The control setup of each unit cell then consists of two speakers, which generate control velocities $v^A_n$ and $v^B_n$, as well as two microphones, which measure the pressure signals $p^A_n$ and $p^B_n$. Based on these measurements, the actuators create the $\gamma$ and $\eta$ couplings in real time. 
The structure of one actuator is detailed in Fig. \ref{fig:speaker}(a). This is an electrodynamic loudspeaker within a closed cavity, which features a diaphragm mechanically driven by a voice coil, placed in a permanent magnetic field. At low frequencies, the loudspeaker can be approximated as a mass-spring-damper system, and the diaphragm motion at small displacements can be described in the Laplace domain by \cite{rivet2016broadband,geib2021tunable}
\begin{subequations} \label{eq:speaker_eqs}
\begin{align}
    Z_{mo}(s)v(s)&=-S_dp(s)+Bli(s), \\
 u(s)&=Z_{eb}(s)i(s)+Blv(s).
\end{align} 
\end{subequations}
Here, $Z_{mo}(s)=M_{ms}s+R_{ms}+\frac{1}{C_{ms}s}$ and $Z_{eb}(s)=L_es+R_e$ are, respectively, the open circuit mechanical and the blocked electrical impedance of the loudspeaker, where
$M_{ms}$, $R_{ms}$, and $C_{mc}$ represent its moving mass, mechanical damping, and the total mechanical compliance. $S_d$ is the effective area of the diaphragm, with $p$ being the total sound pressure acting on it, which includes both the incident and scattered pressure. $v$ is the vibration velocity of the speaker diaphragm, $i$ is the current in the voice coil, and $u$ is the input voltage between the electrical terminals. $Bl$ is the force factor of the speaker, where $B$ is the magnetic field strength and $l$ is the length of the voice coil wire in the magnetic field. 
$R_e$ is the DC resistance, and $L_e$ is the self-inductance of the voice coil. 
To avoid the impact of the coil inductance $L_e$ on the system stability, we designed our loudspeakers to be driven by current sources. We set the control law in each unit cell to
\begin{equation} \label{eq:i_command}
    i_n^{A/B}(s)=(G_S\pm sG_\gamma)p_n^{A/B}(s)+G_\eta\left[p_n^{B/A}(s)-p_n^{A/B}(s)\right],
\end{equation}
which results in the acoustic velocity commands
\begin{equation} \label{eq:v_command}
\begin{split}
    v_n^{A/B}(s)&=\frac{-S_d+Bl(G_S\pm sG_\gamma)}{Z_{mo}(s)}p_n^{A/B}(s)\\&+\frac{BlG_\eta}{Z_{mo}(s)}\left[p_n^{B/A}(s)-p_n^{A/B}(s)\right].
\end{split}    
\end{equation}
Comparing Eq. \eqref{eq:v_command} with Eq. \eqref{eq:A_B_velocities}, and assuming $Z_{mo}(s)\approx M_{ms}s$ in our working frequency range (above the mechanical resonance frequency of the loudspeakers), the controller gains take the form
\begin{equation}  \label{eq:controllers_i}
G_S=\frac{S_d}{Bl}, \; G_\gamma=\frac{ \gamma M_{ms}}{Blz_0}, \; G_\eta=\frac{(\eta-1)M_{ms}}{\rho_0aBl}.
\end{equation}
The structure of the control law in Eq. \eqref{eq:i_command} with Eq. \eqref{eq:controllers_i} is schematically illustrated in Fig. \ref{fig:speaker}(b).
The gain $G_S$ is responsible for the cancellation of the internal physical feedback of the loudspeaker, whereas $G_\gamma$ and $G_\eta$ create the $\gamma$ and the $\eta$ couplings, respectively. 
We then demonstrate the performance of our controlled waveguide using a numerical experiment with $\beta=0.32$ (to be aligned with the actual experiment, Sec. \ref{Experiment_sec}). We calculate the controller gains for the cases $\eta=1.5,2,3$. For these values of $\eta$, we have $\hat{\eta}=1.08,1.16,1.32$ by Eq. \eqref{eq:balance}, then $\hat{\gamma}=0.06,0.11,0.21$ by Eq. \eqref{eq:gamma_eta_balance}, and $\gamma=0.17,0.34,0.66$ back by Eq. \eqref{eq:balance}. Using final modification, the corresponding values of $\gamma=0.16,0.31,0.60$ are obtained. 
The dispersion relations of the resulting closed loops, calculated at low frequencies by PWE, are depicted in Fig. \ref{fig:speaker}(c), on top of the uncontrolled case $\eta=1,\gamma=0$. The spectrum slope increases with $\gamma$, indicating an increase in group velocity. However, due to $\beta<1$, the achieved slope, and thus the $v_g$ increase for the same $\eta$ is less than in Fig. \ref{fig:schematics_and_dispersion}(e).

A waveguide controlled by the feedback system in Fig. \ref{fig:speaker}(a)-(b) for the above combinations of $\gamma$ and $\eta$ was then simulated in the time domain using FE.
The FE model consisted of a waveguide with four unit cells and was terminated on both ends by absorbing boundary conditions. A low-pass filter was applied to the current control signals to minimize the generation of higher-order non-plane waves in the waveguide.
The responses to a Gaussian wave packet are given in Fig. \ref{fig:speaker}(d). 
The group velocities calculated from these responses are depicted in Fig. \ref{fig:speaker}(e) on top of the theoretical expression in Eq. \eqref{eq:vg_increase} (via the mapping in Eq. \eqref{eq:balance}), and those calculated from the dispersion curves in Fig. \ref{fig:speaker}(c), both for the central frequency of the wavepacket, and zero (the dispersion slope is monotonically decreasing with the frequency). The group velocities obtained from the time domain simulations are closer to the small frequency values.

\section{Experimental demonstration}  \label{Experiment_sec}

To confirm the group velocity increase, an active metastructure was fabricated, as shown in the schematic of Fig. \ref{fig:experiment}(a), and the photograph of Fig. \ref{fig:experiment}(b). 
The sound source is positioned at the left end of the waveguide, whereas the right end contains glass wool for sound wave absorption, sealed by a $4$ $\mathrm{mm}$ thick resin block. The waveguide has a uniform wall thickness of $4$ $\mathrm{mm}$, produced via 3D printing using a resin material with a density of $1180$ $\mathrm{kg/m^3}$. The waveguide has a length of $1$ $\mathrm{m}$ and a square cross-section with dimensions of $15 \times 15$ $\mathrm{mm}^2$, corresponding to a cutoff frequency for the plane wave mode of $11467$ $\mathrm{Hz}$. The active metamaterial, $20$ $\mathrm{cm}$ in total length, consists of four unit cells ($8$ active sites) positioned in the center of the waveguide. 

Each unit cell, measuring $a=5$ $\mathrm{cm}$, contains two identical loudspeakers on the top, and two identical microphones on the bottom. 
Two additional microphones, Mic1 and Mic4, are symmetrically positioned on the left and right sides of the active metamaterial, spaced $60$ $\mathrm{cm}$ apart, to monitor the velocity of acoustic wave propagation. 
A controller corresponding to Fig. \ref{fig:speaker}(b) was designed to regulate the output of the control sources $A$ and $B$. This was realized using voltage-controlled current sources (VCCS), which convert a voltage signal into a proportional current signal with unity gain to drive the loudspeakers, as detailed in App. \ref{Controller_VCCS}.
The output voltage signals of the controller, $u^A_{n}$ and $u^B_{n}$, are therefore converted into current signals as $i^{A/B}_n=Gu^{A/B}_n$, where $G=1$ $\mathrm{A/V}$.

To ensure consistency between the loudspeakers and microphones, careful selections and calibrations were performed. The Thiele \& Small parameters of the loudspeakers were measured using the Klippel electroacoustic test system. 
The following average parameters were obtained: the diaphragm mass $M_{ms}=0.0689$ $\mathrm{g}$, the force factor $Bl=0.606$ $\mathrm{N/A}$, the effective diaphragm area $S_d=72$ $\mathrm{mm^2}$, leading to $\beta=0.32$, and the resonance frequency $900$ $\mathrm{Hz}$. 
Electret condenser microphones were used as the pressure sensors. The microphone signals were amplified to ensure a sensitivity of $S_0=400$ $\mathrm{mV/Pa}$. The controller was designed using analog circuitry and fabricated using a printed circuit board, see App. \ref{Controller_VCCS}. Potentiometers were used to adjust $G_\gamma$ and $G_\eta$ to achieve the desired different values of $\gamma$ and $\eta$ according to Eq. \eqref{eq:controllers_i}, divided by $S_0$.
An NI acquisition card with a sampling rate of $f_s=200$ $\mathrm{kHz}$ was used to record the microphone signals from Mic1 and Mic4.

The signal supplied to the sound source was a Gaussian pulse modulated by a $2500$ $\mathrm{Hz}$ sine wave.
Three sets of measurements were conducted with $\eta=1$, $1.5$, and $2$, respectively, and the corresponding $\gamma$ was calculated from Eqs. \eqref{eq:balance} and \eqref{eq:gamma_eta_balance} to be $0$, $0.16$, and $0.31$. 
The measurement results are presented in Fig. \ref{fig:experiment}(c), with a close-up in Fig. \ref{fig:experiment}(d). These results compare the pulse waveform over time for the Hermitian case $\gamma=0$, $\eta=1$ with the non-Hermitian cases $\eta=1.5$ and $\eta=2$. For higher $\eta$ we observed instability in the measured responses.
Setting the peak time of the pulse recorded by microphone Mic1 as $0$, the peak arrival times at microphone Mic4 are advanced by $\delta_t=0.01$ $\mathrm{ms}$ and $\delta_t=0.02$ $\mathrm{ms}$ for $\eta=1.5$ and $\eta=2$, respectively. 
This confirms that the acoustic wave speed was increased.

The actual group velocity $v_g$ of the sound wave in the metamaterial can be determined by the equation $\frac{l}{c}-\frac{l}{v_g}=\delta_t$, where $l=20$ $\mathrm{cm}$ is the total length of the active part. The speed $c$ was obtained as $333$ $\mathrm{m/s}$ by measurement, i.e., $c=s/t_0$, where $s=60$ $\mathrm{cm}$ denotes the distance between microphones Mic1 and Mic4, and $t_0=1.8$ $\mathrm{ms}$ is the time difference between the pulse arriving at Mic4 and Mic1 in the Hermitian case. 
Therefore, for the non-Hermitian cases $\eta=1.5$ and $\eta=2$, the sound wave velocities in the active metamaterial read $339$ $\mathrm{m/s}$ and $345$ $\mathrm{m/s}$, respectively. 
Fig. \ref{fig:experiment}(e) compares experimental and theoretical results (FE time domain of Fig. \ref{fig:speaker}(d)).
The measured Hermitian velocity $c=333$ $\mathrm{m/s}$ is less than the actual sound velocity due to the lower temperature of our experimental environment. 
Correcting to the room temperature sound speed $c=343$ $\mathrm{m/s}$, the experimental results become $v_g=349$ $\mathrm{m/s}$ for $\eta=1.5$ and $v_g=355$ $\mathrm{m/s}$ for $\eta=2$. The corrected values, shown as the solid line in Fig. \ref{fig:experiment}(e), align closely with the theoretical results. 


\section{Discussion and Conclusion}

The balance between the local gain-loss couplings $\gamma$ and the non-local couplings $\eta$ that we derived for the waveguide system via the analogous lattice system, Eqs. \eqref{eq:balance} and \eqref{eq:gamma_eta_balance}, addresses the challenge of merging a spatially discrete coupling pattern into the continuous waveguide medium. The final modification of the balance due to the analogy being valid only in the long wavelength regime, was minor for small $\eta$, but still required.
As a result, a real spectrum was restored in the low frequency range and the desired faster-than-sound dynamics was achieved, 
manifested by the group velocity growth proportional to the square root of $\gamma$. 

FE and PWE simulations suggested that, theoretically, the waveguide system was stable for the achieved group velocities. 
In our experiment, we managed to maintain stable propagation for a velocity increase of up to $355$ $\mathrm{m/s}$ from the nominal $343$ $\mathrm{m/s}$. We anticipate that this limit can be further pushed by optimizations of the control setup and algorithm. 
Our active coupling approach is advantageous in applications that forbid waveguide cross-section blocking, and/or require long-range coupling between multiple active cells. Compared to purely passive systems, the active approach 
enables the modulation profile to be easily reconfigured to more complex interactions, which could lead to new wave properties and wave-guiding capabilities.

\section*{Acknowledgements}

\textit{K. W., H. Z., and J. L. have been supported by the National Natural Science Foundation of China (grant no. 12274221). F. L. has been partially supported by the UK's Engineering and Physical Sciences Research Council (EPSRC) through the 3rd funding call by the UK Acoustics Network Plus EP/V007866/1. C.S. acknowledges support from the National Science Foundation under Grant No. ECCS-2337069. L.S. was partially supported by the Israel Science Foundation Grants No. 2177/23 and 2876/23.}

\appendix

\renewcommand{\thefigure}{A\arabic{figure}}
\renewcommand{\theequation}{A\arabic{equation}}
\setcounter{equation}{0}
\setcounter{figure}{0}

\section{Mapping between the lattice and the waveguide}  \label{cont_discrete_map}

\begin{figure}[tb]
    \centering
    \begin{tabular}{cc}
        \textbf{(a)}   &     \textbf{(b)} \\     \includegraphics[height=3.3 cm, valign=t]{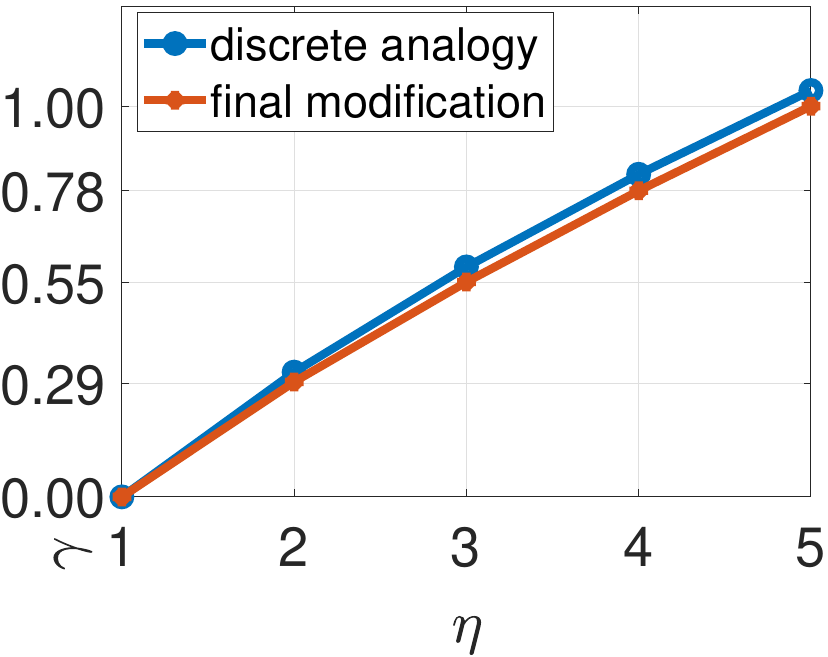}    &          \includegraphics[height=3.3 cm, valign=t]{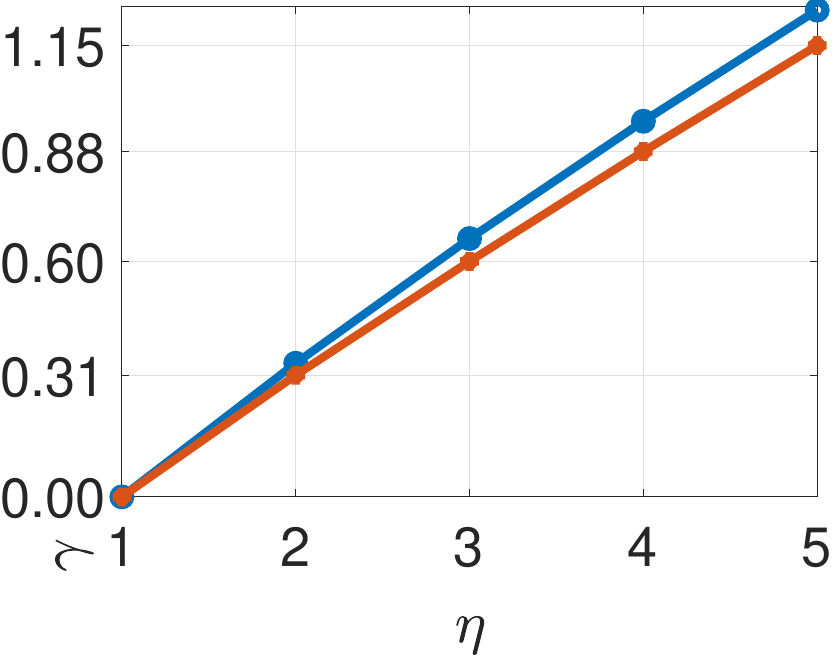}
    \end{tabular}    
    \caption{The PT-symmetry $\gamma-\eta$ balance in the waveguide. Blue - the $\gamma$ value implied by Eqs. \eqref{eq:balance} and \eqref{eq:gamma_eta_balance}. Orange - the actual $\gamma$ value obtained by final modification. (a) $\beta=1$. (b) $\beta=0.32$.}
    \label{fig:gamma_eta_final_mod}
\end{figure}

Here we provide derivation details of the mapping in Eq. \eqref{eq:balance}. 
The lattice equation in Eq. \eqref{eq:lattice_n_eq} is the result of the inhomogeneous mass-spring system
\begin{equation}  \label{eq:lattice_OL_F}
    \begin{cases}
        M_0\Ddot{y}^A_n=K_0\left(y^B_n-2y^A_n+y^B_{n-1}\right)+F^A_n, \\
        M_0\Ddot{y}^B_n=K_0\left(y^A_{n+1}-2y^B_n+y^A_n\right)+F^B_n,
    \end{cases}
\end{equation}
being controlled in closed loop by the input forces
\begin{equation} \label{eq:lattice_F}
    \begin{cases} F_n^A=K_0\left(\hat{\eta}-1\right)\left[y^B_n-y^A_n\right]+\hat{\gamma}\sqrt{K_0M_0} \dot{y}^A_n, \\ F_n^B=K_0\left(\hat{\eta}-1\right)\left[y^A_n-y^B_n\right]-\hat{\gamma}\sqrt{K_0M_0} \dot{y}^B_n. \end{cases}
\end{equation}
Since $M_0\ddot{y}_n=K_0\left(y_{n+1}-2y_n+y_{n-1}\right)+F_n$, a general forced lattice equation,
maps to the inhomogeneous acoustic waveguide equation \eqref{eq:cont_OL}, and the lattice site displacement $y$ maps to the integral of pressure $p$, the lattice control inputs $F$ in Eq. \eqref{eq:lattice_F} are mapped to the acoustic inputs as $\beta v$, i.e.
\begin{equation}   \label{eq:V_AV_B_correct}
    \begin{cases} v^A_n=\dfrac{2(\hat{\eta}-1)}{\beta\rho_0a}\displaystyle\int_0^t\left[p^B_n-p^A_n\right]\mathrm{d}t+\frac{\hat{\gamma}}{\beta z_0} p^A_n, \\ v^B_n=\dfrac{2(\hat{\eta}-1)}{\beta\rho_0a}\displaystyle\int_0^t\left[p^A_n-p^B_n\right]\mathrm{d}t-\frac{\hat{\gamma}}{\beta z_0} p^B_n.\end{cases} 
\end{equation}
Comparing Eq. \eqref{eq:V_AV_B_correct} with the actual acoustic inputs in Eq. \eqref{eq:A_B_velocities}, the mapping of Eq. \eqref{eq:balance} is straightforwardly obtained. 
The final modification of $\gamma$ for a fixed $\eta$, which is required due to the discrepancy between the continuous and the equivalent discrete model at the crossing point, is illustrated in Fig. \ref{fig:gamma_eta_final_mod}. Both for $\beta=1$ and $\beta=0.32$, the actual value of $\gamma$ that is required to restore PT symmetry in the waveguide is slightly smaller than the value that stems from Eqs. \eqref{eq:balance} and \eqref{eq:gamma_eta_balance}.

\renewcommand{\thefigure}{B\arabic{figure}}
\renewcommand{\theequation}{B\arabic{equation}}
\setcounter{equation}{0}
\setcounter{figure}{0}

\section{Group velocity control in the lattice model}
\label{Lattice}

The analogous lattice model in Fig. \ref{fig:schematics_and_dispersion}(b) is represented by the augmented eigenvalue problem in Eq. \eqref{eq:Hamiltonian}, which reads
\begin{equation}
    \Omega^4-\left(2(1+\hat{\eta})-\hat{\gamma}^2\right)\Omega^2+(1+\hat{\eta})^2-ff^*=0.
\end{equation}
Its solution is given by
\begin{equation} \label{eq:aug_eq}
    \Omega^2=\tfrac{1}{2}\left(2(1+\hat{\eta})-\hat{\gamma}^2\right)\pm\tfrac{1}{2}\sqrt{\delta},
\end{equation}
with $\delta=\left(2(1+\hat{\eta})-\hat{\gamma}^2\right)^2-4\left((1+\hat{\eta})^2-ff^*\right)$.
The requirements for real spectrum (with minimum at $\cos ka=-1$)
give the PT symmetry condition $\hat{\gamma}\leq\hat{\gamma}^*$ of Eq. \eqref{eq:gamma_eta_balance} via Eq. \eqref{eq:balance}. 
We then obtain $(1+\hat{\eta})^2-ff^*=2\hat{\eta}(1-\cos ka)$, $\delta=8\hat{\eta}(1+\cos ka)$, and the dispersion relation is obtained from Eq. \eqref{eq:aug_eq} as
\begin{equation}  \label{eq:Om_discrete}
\Omega=\sqrt{\sqrt{2}\hat{\gamma}^*+2}\cdot\sqrt{1\pm\sqrt{\tfrac{1}{2}\left(1+\cos ka\right)}}.
\end{equation}
In the Hermitian case, the lattice constant is $a/2$. We then use the identity $\cos{ka}=\cos{ka}=\cos^2{\frac{ka}{2}}-\sin^2{\frac{ka}{2}}$, which implies that $\sqrt{\tfrac{1}{2}\left(1+\cos ka\right)}$ equals $\cos{\frac{ka}{2}}$ with the positive branch chosen for $k\in[0,\frac{\pi}{a}]$ and the negative branch for $k\in(\frac{\pi}{a},\frac{2\pi}{a}]$. We then rewrite $\Omega$ in Eq. \eqref{eq:Om_discrete} as a function of the Hermitian frequency spectrum, as
\begin{equation}  \label{eq:Om_c_0}
\Omega=\sqrt{\tfrac{1}{\sqrt{2}}\hat{\gamma}+1}\Omega_H \quad , \quad \Omega_H=\sqrt{2}\sqrt{1-\cos{\tfrac{ka}{2}}}.
\end{equation}
The group velocity (normalized by $\omega_0$) thus becomes
\begin{equation}  \label{eq:v_c_complete}
v_g=\frac{\partial \Omega}{\partial k}=\sqrt{\tfrac{1}{\sqrt{2}}\hat{\gamma}+1}\frac{\partial \Omega_H}{\partial k}=\sqrt{\tfrac{1}{\sqrt{2}}\hat{\gamma}+1}v_{0},
\end{equation}
where $v_{0}$ is the Hermitian group velocity, given by
\begin{equation}
    v_{0}=\frac{a\sin{\tfrac{ka}{2}}}{2\sqrt{2}\sqrt{1-\cos{\tfrac{ka}{2}}}}.
\end{equation}

\renewcommand{\thefigure}{C\arabic{figure}}
\renewcommand{\theequation}{C\arabic{equation}}
\setcounter{equation}{0}
\setcounter{figure}{0}

\section{The Plane Wave Expansion derivation}
\label{PWE_details}

We substitute the proposed solution $p(x,t)=\mathrm{e}^{\mathrm{i}\omega t}P(x)$, with $P(x)=\sum_{m=-M}^M\mathrm{e}^{-\mathrm{i}(k+m)\textbf{b}_1\cdot x\textbf{d}_1}\overline{p}_m$, $\textbf{b}_1=\frac{2\pi}{a}\hat{\textbf{e}}_1$ and $\textbf{d}_1=a\hat{\textbf{e}}_1$, into the $n_{th}$ unit cell of Eq. \eqref{eq:cont}. Using a general formulation $\textbf{r}=x\textbf{d}_1$, $\textbf{k}=k\textbf{b}_1$, and $\textbf{G}=m\textbf{b}_1$, and transforming $p_{tt},p_t$ to frequency domain as $-\omega^2p,\mathrm{i}\omega p$, reads
\begin{equation}  \label{eq:PWE_sol_subs}   \begin{split}
    &\frac{\omega^2}{c^2}\sum_{m=-M}^Me^{-\mathrm{i}(\textbf{k}+\textbf{G})\cdot \textbf{r}}\overline{p}_m=\sum_{m=-M}^M |\textbf{k}+\textbf{G}|^2e^{-\mathrm{i}(\textbf{k}+\textbf{G})\cdot \textbf{r}}\overline{p}_m\\&-\beta\sum_{m=-M}^M \big[\frac{\eta-1}{a}\left(e^{-\mathrm{i}(\textbf{k}+\textbf{G})\cdot \textbf{R}_B}-e^{-\mathrm{i}(\textbf{k}+\textbf{G})\cdot \textbf{R}_A}\right)\\&-\frac{\gamma}{c} \mathrm{i}\omega e^{-\mathrm{i}(\textbf{k}+\textbf{G})\cdot \textbf{R}_A}\big]\delta(\textbf{r}-\textbf{R}_A)\overline{p}_m\\&-\beta\sum_{m=-M}^M \big[\frac{\eta-1}{a}\left(e^{-\mathrm{i}(\textbf{k}+\textbf{G})\cdot \textbf{R}_A}-e^{-\mathrm{i}(\textbf{k}+\textbf{G})\cdot \textbf{R}_B}\right)\\&+\frac{\gamma}{c} \mathrm{i}\omega e^{-\mathrm{i}(\textbf{k}+\textbf{G})\cdot \textbf{R}_B}\big]\delta(\textbf{r}-\textbf{R}_B)\overline{p}_m,
    \end{split}
\end{equation}
where $\textbf{R}_A=-\frac{1}{4}\textbf{d}_1$ and $\textbf{R}_B=\frac{1}{4}\textbf{d}_1$ are the actuators locations measured from the center of the unit cell. Multiplying Eq. \eqref{eq:PWE_sol_subs} by $e^{\mathrm{i}(\textbf{k}+\widehat{\textbf{G}})\cdot \textbf{r}}$, where $\widehat{\textbf{G}}=\widehat{m}\textbf{b}_1$, gives
\begin{equation}  \label{eq:PWE_sol_subs_mhat}   
\begin{split}
    &\frac{\omega^2}{c^2}\sum_{m=-M}^Me^{-\mathrm{i}(\textbf{G}-\widehat{\textbf{G}})\cdot \textbf{r}}\overline{p}_m=\sum_{m=-M}^M |\textbf{k}+\textbf{G}|^2e^{-\mathrm{i}(\textbf{G}-\widehat{\textbf{G}})\cdot \textbf{r}}\overline{p}_m\\&-\beta\sum_{m=-M}^Me^{\mathrm{i}(\textbf{k}+\widehat{\textbf{G}})\cdot \textbf{r}}\Big(\frac{\eta-1}{a}\left[e^{-\mathrm{i}(\textbf{k}+\textbf{G})\cdot \textbf{R}_B}-e^{-\mathrm{i}(\textbf{k}+\textbf{G})\cdot \textbf{R}_A}\right]\\&-\frac{\gamma}{c} \mathrm{i}\omega e^{-\mathrm{i}(\textbf{k}+\textbf{G})\cdot \textbf{R}_A}\Big)\delta(\textbf{r}-\textbf{R}_A)\overline{p}_m\\&-\beta\sum_{m=-M}^M e^{\mathrm{i}(\textbf{k}+\widehat{\textbf{G}})\cdot \textbf{r}}\Big(\frac{\eta-1}{a}\left[e^{-\mathrm{i}(\textbf{k}+\textbf{G})\cdot \textbf{R}_A}-e^{-\mathrm{i}(\textbf{k}+\textbf{G})\cdot \textbf{R}_B}\right]\\&+\frac{\gamma}{c} \mathrm{i}\omega e^{-\mathrm{i}(\textbf{k}+\textbf{G})\cdot \textbf{R}_B}\Big)\delta(\textbf{r}-\textbf{R}_B)\overline{p}_m.
    \end{split}
\end{equation}
Due to the orthogonality of the Fourier series, we have
\begin{equation}  \label{eq:PWE_int_terms}
    \int_{A_c}e^{-\mathrm{i}(\textbf{G}-\widehat{\textbf{G}})\cdot \textbf{r}}\mathrm{d}A_c=\begin{cases}A_c, \quad  &\textbf{G}=\widehat{\textbf{G}} \\ 0, \quad &\textbf{G}\neq\widehat{\textbf{G}}\end{cases} .
\end{equation}
Also we have
\begin{equation}    \int_{A_c}f(\textbf{r})\delta(\textbf{r}-\textbf{R}_\alpha)\mathrm{d}A_c=f(\textbf{R}_\alpha),
\end{equation}
where $A_c=a$ is the unit cell length and the lattice constant. Integrating Eq. \eqref{eq:PWE_sol_subs_mhat} over a unit cell, and solving for all $\widehat{m}$ in the range $-M,...,M$, then gives
\begin{equation}  \label{eq:after_orth}
\begin{split}
   \frac{\omega^2}{c^2}a\textbf{I}_N\overline{\textbf{p}}&=\textbf{W}a\overline{\textbf{p}}\\&-\beta\sum_{m=-M}^M\Big(\frac{\eta-1}{a}e^{\mathrm{i}\textbf{k}\cdot(\textbf{R}_A-\textbf{R}_B)}e^{\mathrm{i}\widehat{\textbf{G}}\cdot \textbf{R}_A}e^{-\mathrm{i}\textbf{G}\cdot \textbf{R}_B}\overline{p}_m\\&+\left[\frac{\eta-1}{a}+\frac{\gamma}{c} \mathrm{i}\omega\right]e^{-\mathrm{i}(\textbf{G}-\widehat{\textbf{G}})\cdot \textbf{R}_A}\overline{p}_m\Big)\\
    &-\beta\sum_{m=-M}^M\Big(\frac{\eta-1}{a}e^{\mathrm{i}\textbf{k}\cdot(\textbf{R}_B-\textbf{R}_A)}e^{\mathrm{i}\widehat{\textbf{G}}\cdot \textbf{R}_B}e^{-\mathrm{i}\textbf{G}\cdot \textbf{R}_A}\overline{p}_m\\&+\left[\frac{\eta-1}{a}-\frac{\gamma}{c} \mathrm{i}\omega\right]e^{-\mathrm{i}(\textbf{G}-\widehat{\textbf{G}})\cdot \textbf{R}_B}\overline{p}_m\Big),
    \end{split}
\end{equation}
where $N=2M+1$ is the total number of terms in the series,
\begin{equation}
    \textbf{W}=\left(\begin{array}{ccc} |\textbf{k}+\widehat{\textbf{G}}_1|^2 &  &  \\
          & \cdots & \\
          & & |\textbf{k}+\widehat{\textbf{G}}_N|^2
    \end{array}\right),
\end{equation}
$\overline{\textbf{p}}$ is the eigenvector of length $N$, $\textbf{G}\cdot\textbf{R}_A=-\frac{\pi}{2}m$, and $\textbf{G}\cdot\textbf{R}_B=\frac{\pi}{2}m$.
Using matrix formulation, we define $\sum_{m=-M}^M e^{\mathrm{i}\widehat{\textbf{G}}\cdot \textbf{R}_{\widehat{j}}}e^{-\mathrm{i}\textbf{G}\cdot \textbf{R}_j}\overline{p}_m=\textbf{E}_{\widehat{j}j} \overline{\textbf{p}}$, where
\begin{equation}  \label{eq:E_j}
    \textbf{E}_{\widehat{j}j}=e^{\mathrm{i}\left[\begin{array}{cccc}{\textbf{G}_1}
         & \textbf{G}_2 & \cdots & \textbf{G}_N\end{array}\right]'\cdot \textbf{R}_{\widehat{j}}}\cdot e^{-\mathrm{i}\textbf{R}_j\cdot\left[\begin{array}{cccc}{\textbf{G}_1}
         & \textbf{G}_2 & \cdots & \textbf{G}_N\end{array}\right]},
\end{equation}
and
$j$ and $\widehat{j}$ take the according values of $A$ and $B$. Eq. \eqref{eq:after_orth} then takes the form of the polynomial eigenvalue problem
\begin{equation}  \label{eq:poly_eig}    \left(\textbf{q}_2\omega^2+\textbf{q}_1\omega+\textbf{q}_0\right)\overline{\textbf{p}}=0,
\end{equation}
where
\begin{equation}  \label{eq:poly_eig_ABC_MC_K}
\begin{split}
    \textbf{q}_2&=\frac{a}{c^2}\textbf{I}_N, \\ \textbf{q}_1&=-\beta\frac{\gamma}{c} \mathrm{i}\left(\textbf{E}_{\widehat{A}A}-\textbf{E}_{\widehat{B}B}\right), \\ \textbf{q}_0&=-a\textbf{W}+\beta \frac{\eta-1}{a}\Big(e^{\mathrm{i}\textbf{k}\cdot(\textbf{R}_A-\textbf{R}_B)}\textbf{E}_{\widehat{A}B}+\textbf{E}_{\widehat{A}A}\\&+e^{\mathrm{i}\textbf{k}\cdot(\textbf{R}_B-\textbf{R}_A)}\textbf{E}_{\widehat{B}A}-\textbf{E}_{\widehat{B}B}\Big).
    \end{split}
\end{equation}
We then rewrite Eqs. \eqref{eq:poly_eig}-\eqref{eq:poly_eig_ABC_MC_K} in a companion form to obtain an augmented linear eigenvalue problem
\begin{equation}  \label{eq:lin_eig}
    \omega\textbf{P}\textbf{v}=\textbf{Q}\textbf{v},
\end{equation}
where $\textbf{v}$ is the augmented eigenvector of length $2N$, and
\begin{equation}  \label{eq:lin_eig_mat}
    \textbf{P}=\left(\begin{array}{cc} \textbf{I} & 0 \\
         0 & \textbf{q}_2
    \end{array}\right), \quad \textbf{Q}=\left(\begin{array}{cc} 
         0 & \textbf{I} \\
        -\textbf{q}_0 & -\textbf{q}_1 
    \end{array}\right).
\end{equation}
The solution of Eqs. \eqref{eq:lin_eig}-\eqref{eq:lin_eig_mat} gives the black and orange dispersion curves in Fig. \ref{fig:schematics_and_dispersion}(c)-(f).

\renewcommand{\thefigure}{D\arabic{figure}}
\renewcommand{\theequation}{D\arabic{equation}}
\setcounter{equation}{0}
\setcounter{figure}{0}

\section{The controller structure}
\label{Controller_VCCS}

\begin{figure}[tb]
    \centering
    \begin{tabular}{cc}
    & \\
        \textbf{(a)}   &          \includegraphics[height=4.5 cm, valign=t]{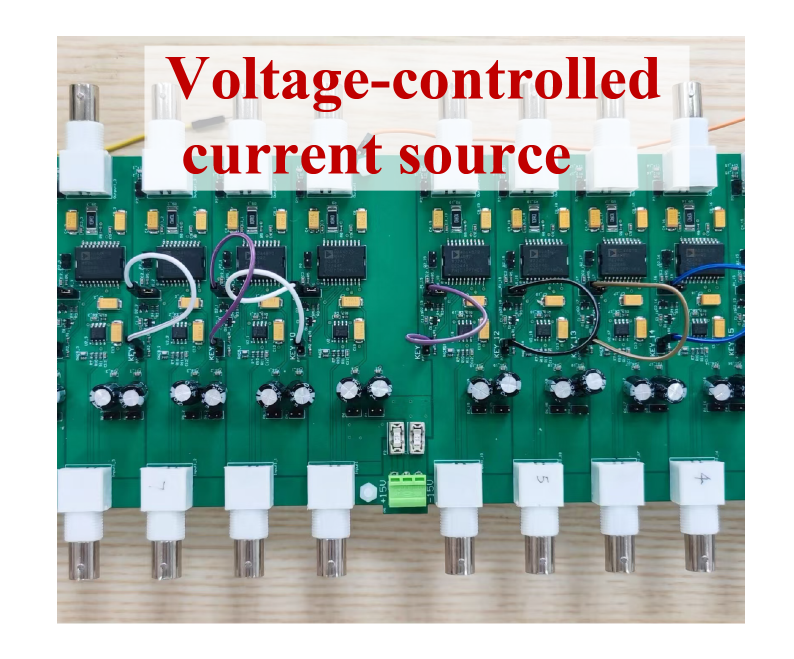} \\
        \textbf{(b)}   &          \includegraphics[height=4.1 cm, valign=t]{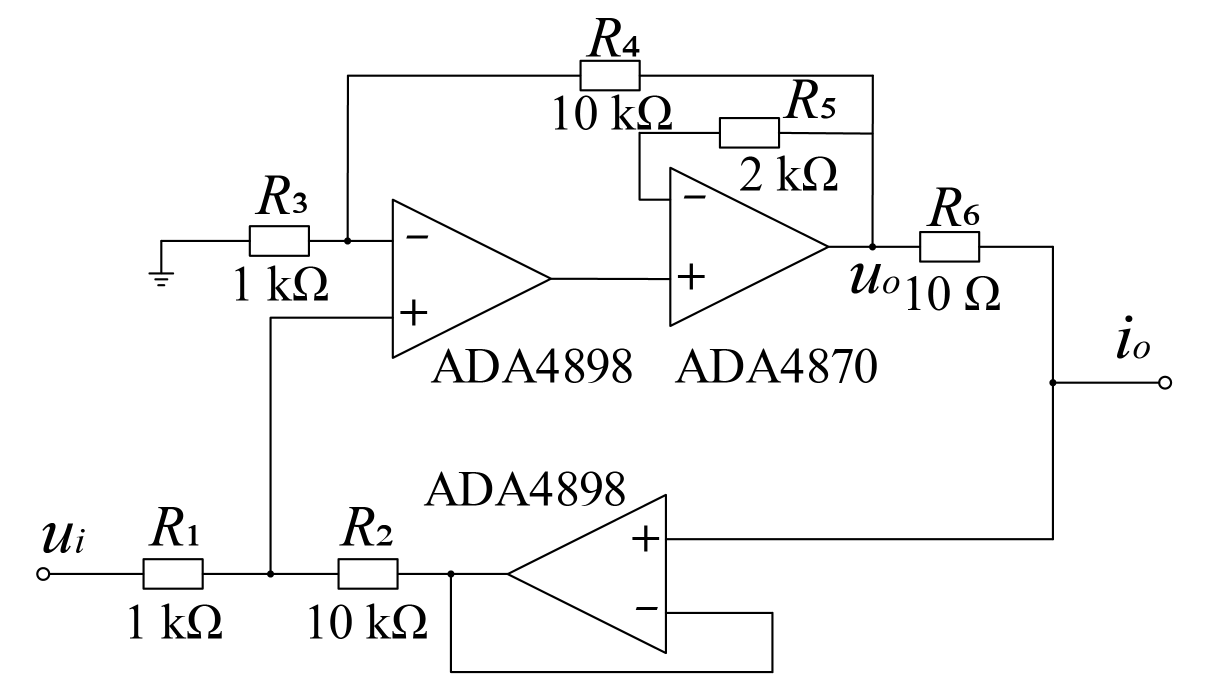} \\
        \textbf{(c)}   &          \includegraphics[height=4.5 cm, valign=t]{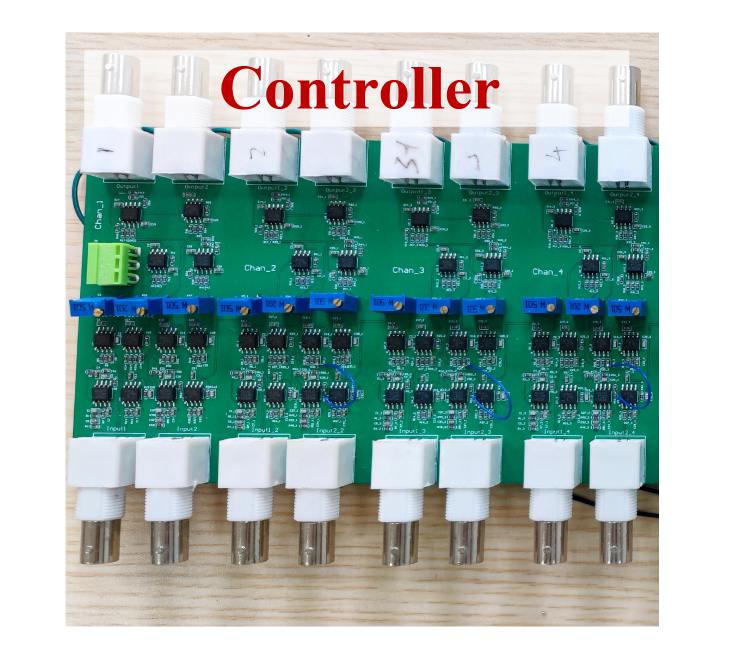}
    \end{tabular}    
    \caption{Realization of the controller of Fig. \ref{fig:speaker}(b). (a),(b) Photograph and circuitry of the Voltage-Controlled Current Sources. (c) Photograph of the controller.}
    \label{fig:VCCS_scheme}
\end{figure}

The VCCS employed in our experiment, photographed in Fig. \ref{fig:VCCS_scheme}(a), were designed using an improved Howland current pump circuit \cite{lam2023analysis}. 
It integrates the high output of the operational amplifier ADA4870 with the fast response of the operational amplifier ADA4898. 
The relation between the output current $i_o$ and the input voltage $u_i$ then becomes $i_o=\frac{R_2}{R_1R_6}u_i+    \frac{R_2R_3-R_1R_4}{R_1R_6(R_3+R_4)}u_o$, as illustrated in Fig. \ref{fig:VCCS_scheme}(b).
Substituting the resistor values from Fig. \ref{fig:VCCS_scheme}(b) yields the output current $i_o=Gu_i$, where $G=1$ $\mathrm{A/V}$. The circuit performance was verified via Multisim simulations. The resulting controller is photographed in Fig. \ref{fig:VCCS_scheme}(c).

\bibliographystyle{IEEEtran}

\bibliography{fastwavepackets}

\end{document}